\documentclass[twocolumn]{autart}    

\usepackage{amsmath,amssymb,amsfonts}
\usepackage{algorithm,algorithmic}
\usepackage{graphicx}
\usepackage{textcomp}
\usepackage{url}
\usepackage{color}

\usepackage{lipsum}
\usepackage{epstopdf}

\usepackage{amsopn}
\DeclareMathOperator{\diag}{diag}

\usepackage{enumerate}
\usepackage{cancel}
\usepackage{mathrsfs}
\usepackage{mathdots}
\usepackage{euscript}
\usepackage{amscd}
\usepackage{placeins}
\usepackage{natbib}

\newtheorem{theorem}{Theorem}
\newtheorem{corollary}{Corollary}
\newtheorem{lemma}{Lemma}
\newtheorem{proposition}{Proposition}

\theoremstyle{definition}
\newtheorem{definition}{Definition}
\newtheorem{example}{Example}

\theoremstyle{remark}
\newtheorem{remark}{Remark}

\newcommand{\bmat}{\left[ \begin{matrix}}
	\newcommand{\emat}{\end{matrix} \right]}
\newcommand{\innerprod}[2]{\langle{#1},\,{#2}\rangle}

\DeclareMathOperator{\trace}{tr}
\DeclareMathOperator{\conv}{conv}

\DeclareMathOperator{\E}{{\mathbb E}}
\newcommand{\Rbb}{\mathbb R}

\newcommand{\Cbb}{\mathbb C}

\newcommand{\Zbb}{\mathbb Z}

\newcommand{\Tbb}{\mathbb T}

\newcommand{\xb}{\mathbf  x}
\newcommand{\yb}{\mathbf  y}
\newcommand{\sbf}{\mathbf  s}  

\newcommand{\gb}{\mathbf  g}

\newcommand{\ab}{\mathbf a}

\newcommand{\tb}{\mathbf t}
\newcommand{\kb}{\mathbf k}

\newcommand{\zerob}{\mathbf 0}

\newcommand{\Ab}{\mathbf A}

\newcommand{\Gb}{\mathbf G}

\newcommand{\Pb}{\mathbf P}
\newcommand{\Qb}{\mathbf Q}

\newcommand{\Ub}{\mathbf U}

\newcommand{\Xb}{\mathbf  X}

\newcommand{\thetab}{\boldsymbol{\theta}}

\newcommand{\sigmab}{\boldsymbol{\sigma}}

\newcommand{\Sigmab}{\boldsymbol{\Sigma}}

\newcommand{\Lambdab}{\boldsymbol{\Lambda}}

\DeclareMathOperator{\range}{Range}
\DeclareMathOperator{\rank}{rank}

\newcommand{\Cfrak}{\mathfrak{C}}
\newcommand{\Pfrak}{\mathfrak{P}}

\newcommand{\Acal}{\mathcal{A}}

\newcommand{\Zcal}{\mathcal{Z}}

\renewcommand{\d}{\mathrm{d}}

\newcommand{\SNR}{\mathrm{SNR}}

\newcommand{\trm}{\mathrm{t}}
\newcommand{\srm}{\mathrm{s}}

\begin{document}

\begin{frontmatter}

\title{Line Spectrum Representation for Vector Processes with Application to Frequency Estimation \thanksref{footnoteinfo}} 

\thanks[footnoteinfo]{This work was supported in part by the ``Hundred-Talent Program'' of Sun Yat-sen University and the National Natural Science Foundation of China under the grant number 62103453. Corresponding author B.~Zhu. Tel. +86 14748797525. Fax +86(20) 39336557.}

\author[SYSU]{Bin Zhu}\ead{zhub26@mail.sysu.edu.cn}    

\address[SYSU]{School of Intelligent Systems Engineering, Sun Yat-sen University, Waihuan East Road 132, 510006 Guangzhou, China}


\begin{abstract}                          

A positive semidefinite Toeplitz matrix, which often arises as the finite covariance matrix of a stationary random process, can be decomposed as the sum of a nonnegative multiple of the identity corresponding to a white noise, and a singular term corresponding to a purely deterministic process. Moreover, the singular nonnegative Toeplitz matrix admits a unique characterization in terms of spectral lines which are associated to an oscillatory signal. This is the content of the famous Carath\'{e}odory-Fej\'{e}r theorem. Its importance lies in the practice of extracting the signal component from noise, providing insights in modeling, filtering, and estimation. The multivariate counterpart of the theorem concerning block-Toeplitz matrices is less well understood, and in this paper, we aim to partially address this issue. To this end, we first establish an existence result of the line spectrum representation for a finite covariance multisequence of some underlying random vector field. Then, we give a sufficient condition for the uniqueness of the representation, which indeed holds true in the special case of bivariate time series. Equivalently, we obtain the Vandermonde decomposition for positive semidefinite block-Toeplitz matrices with $2\times 2$ blocks. The theory is applied to the problem of frequency estimation with two measurement channels within the recently developed framework of atomic norm minimization. It is shown that exact frequency recovery can be guaranteed in the noiseless case under suitable conditions, while in the noisy case, extensive numerical simulations are performed showing that the method performs well in a wide range of signal-to-noise ratios.
\end{abstract}

\end{frontmatter}

\section{Introduction}\label{sec:intro}



Line spectrum analysis has been an important subject of research in science and engineering, and has produced a large body of literature \cite[see e.g., the survey paper][]{stoica1993list}. It is intimately related to the frequency estimation problem of an oscillatory time series and its spatial variant known as Direction-of-Arrival (DOA) estimation in radar signal processing, and they find applications in telecommunication, astronomy, econometrics, and mechanics, among others \citep[cf.~][]{quinn2001estimation,van2004optimum}.


Given a stationary time series, a fundamental question in identification and signal processing is to seek a decomposition of the series into the ``signal-plus-noise'' form.
This leads to the additive decomposition of a positive semidefinite Toeplitz matrix (formed by a finite number of covariances of a stationary scalar process)
\begin{equation}\label{additive_decomp_Toeplitz}
T = T_\srm + \alpha I,
\end{equation}
in which the signal part $T_\srm\geq 0$ is singular and the noise part has the largest possible variance/energy $\alpha\geq0$. Such a decomposition is canonical in the sense that there is a \emph{unique} power spectrum, consisting of \emph{spectral lines}, consistent with the covariance data in $T_\srm$. This uniqueness result is due to Carath\'{e}odory and Fej\'{e}r \citep[see][]{Grenander_Szego} and rediscovered later in \cite{pisarenko1973retrieval} which forms the foundation of modern subspace methods such as MUSIC and ESPRIT \citep[cf.~][]{stoica2005spectral}. More recently, the Carath\'{e}odory-Fej\'{e}r decomposition (also called \emph{Vandermonde decomposition}, a name that will be adopted in later sections) serves as a key technical tool for a class of frequency estimation methods inspired by the compressive sensing literature \citep[see e.g.,][]{candes2014towards,tang2013compressed,yang2018sparse}. A number of generalizations of the Carath\'{e}odory-Fej\'{e}r decomposition exist for {\em scalar} processes. \citet{georgiou2000signal,Georgiou-01} concerns state covariance matrices arising from the output of a filter bank which contain positive semidefinite Toeplitz matrices as a special case, while \cite{yang2018frequency} study a frequency-selective\footnote{The term ``frequency-selective'' means that the spectral lines are located in some \emph{a priori} known subinterval of $(-\pi,\pi]$. In this paper however, we study the usual Vandermonde decomposition for \emph{vector} processes without imposing such an additional restriction.} version of the decomposition. \citet{lang1983spectral} focus on the line spectrum representation and Pisarenko's estimate for scalar random fields, and the papers
\cite{yang2016vandermonde,andersson2017structure} deal with a generalization of the Vandermonde decomposition to multilevel Toeplitz matrices.

Singularity in the covariance matrix reveals the linear dependence between the values of the time series, and such a process is termed \emph{purely deterministic} \citep{LP15}. The same observation can be made in the multivariate case, i.e., for stationary vector processes. However, the corresponding results of the Carath\'{e}odory-Fej\'{e}r type are much fewer.
We mention the important and rather technical work \cite{georgiou2007caratheodory} which studies the corresponding question of additive decomposition for block-Toeplitz matrices. It is claimed in that paper that a canonical decomposition similar to \eqref{additive_decomp_Toeplitz}, where the signal component has a unique line spectrum representation, in general does not exist. 
As a complement to that result, we show in the present work that at least in the bivariate case, a singular positive semidefinite block-Toeplitz matrix can indeed be characterized by spectral lines in a unique fashion under a mild condition of nondegeneracy for the matricial covariance sequence.

The roadmap and the contributions of this paper are briefly described next. We begin by formulating the problem of line spectrum representation for stationary random vector fields\footnote{Stationary processes over the integer grid are also known as homogeneous random fields.} and reviewing the conic characterization of the extendability of a covariance multisequence. The formulation can be viewed as an instance of a ``moment problem'' which is reminiscent of the vast literature on \emph{covariance extension}, see e.g., \citet{byrnes2000anewapproach,Georgiou-06,karlsson2013uncertainty,RKL-16multidimensional,zhu2020well,ZFKZ2019M2,zhu2020m} and the references therein. Then we make use of techniques from measure theory to establish a general existence result on the line spectrum representation for an extendable covariance multisequence, with potentially many impulses. Next, we consider the Carath\'{e}odory-Fej\'{e}r-type decomposition for covariances on the boundary of the dual cone (i.e., the set of extendable covariances) so that the locations of the spectral lines can be limited to the zero set of some nonnegative polynomial. We give a sufficient condition for the uniqueness of the line spectrum representation, which is often too strong to hold in the multidimensional case.
However, when specializing the theory to the particular $1$-d case of time series, we obtain without much difficulty a corollary stating the existence of the Vandermonde decomposition of block-Toeplitz matrices with $2\times 2$ blocks under a nondegeneracy condition.
Moreover, a computational procedure to find such a decomposition is outlined.
Finally, as an illustration of possible applications, we integrate the Vandermonde decomposition of block-Toeplitz matrices into the framework of \emph{atomic norm minimization} for the problem of frequency estimation which has been extensively developed in recent years. Under appropriate conditions, the unknown frequencies can be exactly recovered in the noiseless case. The method can also be extended to the noisy case and its performance is illustrated via extensive numerical simulations.

This paper is organized as follows. Section~\ref{sec:problem} gives the problem formulation in the language of the trigonometric moment problem.
Section~\ref{sec:existence} is dedicated to the existence of a line spectrum representation for a finite covariance multisequence of some underlying random vector field.
Section~\ref{sec:uniqueness} concerns the uniqueness of the representation, yielding in particular the Vandermonde decomposition of covariance matrices of stationary purely deterministic bivariate time series.
Section~\ref{sec:freq_est} presents an application of the Vandermonde decomposition and gives a convex optimization method to retrieve the frequencies from two measurement channels of an oscillatory signal. Section~\ref{sec:conclusions} makes some concluding remarks.


%

\section{Problem formulation}
\label{sec:problem}

Given positive integers $m$ and $d$, consider a second-order stationary zero-mean random complex $m$-vector field $\yb(\tb)$ with $\tb=(t_1,t_2,\dots,t_d)\in\Zbb^d$. Here $\Zbb$ is the set of integers.
The (matricial) covariance function of the random field is defined as the expectation $\Sigma_{\kb}:=\E \yb(\tb+\kb)\yb(\tb)^*$ which does not depend on $\tb$ by stationarity. As a simple consequence, the symmetry with respect to the origin $\Sigma_{-\kb}=\Sigma_{\kb}^*$ holds.
Moreover, the covariance field admits a representation \citep{yaglom1957some}
\begin{equation}\label{Sigma_spec}
\Sigma_{\kb} = \int_{\Tbb^d} e^{i\innerprod{\kb}{\thetab}} \d M(\thetab),\quad \kb\in\Zbb^d,
\end{equation}
where $\thetab=(\theta_1,\theta_2,\dots,\theta_d)$ takes valued in $\Tbb^d:=(-\pi,\pi]^d$, $\innerprod{\kb}{\thetab}:=k_1\theta_1+\cdots+k_d\theta_d$ denotes the inner product in $\Rbb^d$, and $\d M(\thetab)$ is an $m\times m$ Hermitian positive semidefinite matrix-valued measure on $\Tbb^d$.

In practice, only a finite set of covariances are available, typically estimated via some averaging scheme from a realization of the random field, and the problem is to infer the spectral measure $\d M$ based on the given second-order statistics. 
To be more precise, consider a matricial multisequence $\Sigmab:=\{\Sigma_{\kb}\}_{\kb\in\Lambda}$ with indices in a finite set $\Lambda\subset\Zbb^d$ that contains $\zerob$ the all-zero index and has the natural symmetry $\kb\in\Lambda \implies -\kb\in\Lambda$.
Notice that such an index set $\Lambda$ necessarily has an odd number of elements.
In the most common situation, $\Lambda$ is a cuboid centered at the origin which amounts to the set $\{-n,\dots,0,\dots,n\}$ in the $1$-d case. 
Then the problem is to find a nonnegative matricial measure $\d M$ that solves the integral equations
\begin{equation}\label{moment_eqns}
\int_{\Tbb^d} e^{i\innerprod{\kb}{\thetab}} \d M(\thetab) = \Sigma_{\kb}\text{ for all } \kb\in\Lambda.
\end{equation}
However, given the finite multisequence $\Sigmab$, the existence of a solution (namely a nonnegative matricial measure $\d M$) to \eqref{moment_eqns} is a highly nontrivial question in the multivariate ($m>1$) multidimensional ($d>1$) setting. When such existence holds true, we shall call the set of covariances $\Sigmab$ \emph{extendable}, a terminology derived from the classical problem of covariance extension \citep{A65moments,KreinNudelman}. According to \cite{Georgiou-06}, the extendability condition can be characterized as follows. Let $e^{i\thetab}:=(e^{i\theta_1},\dots,e^{i\theta_d})$ denote a point on the $d$-torus (which is isomorphic to $\Tbb^d$), and define
\begin{equation}
\begin{split}
\Pfrak_+ := \left\{ P(e^{i\thetab})= \sum_{\kb\in\Lambda} P_\kb e^{-i\innerprod{\kb}{\thetab}} \,:\, P_{-\kb}=P_\kb^*\in\Cbb^{m\times m}, \right. \\ \left. P(e^{i\thetab})\geq 0 \ \forall \thetab\in\Tbb^d \right\}
\end{split}
\end{equation}
as the set of matricial Hermitian trigonometric polynomials with indices in $\Lambda$ that are positive semidefinite on the $d$-torus.
Apparently, any polynomial in $\Pfrak_+$ can be identified as a multisequence $\Pb:=\{P_\kb\}_{\kb\in\Lambda}$. An inner product between such multisequences can be defined as
\begin{equation}\label{inner_prod}
\innerprod{\Sigmab}{\Pb}:= \sum_{\kb\in\Lambda} \trace(\Sigma_{\kb} P_\kb^*)
\end{equation}
which is in fact real-valued due to the symmetry. 
We can then proceed to define the dual cone as
\begin{equation}\label{dual_cone}
\Cfrak_+ := \left\{ \Sigmab=\{\Sigma_\kb\}_{\kb\in\Lambda} \,:\, \innerprod{\Sigmab}{\Pb}\geq 0 \ \forall P\in\Pfrak_+ \right\}.
\end{equation}
By \citet[Proposition 1, p. 1059]{Georgiou-06}, a multisequence $\Sigmab$ is extendable if and only if $\Sigmab\in\Cfrak_+$.

The above characterization of extendability is conceptually clear, but in practice, it does not offer a ``simple'' testable criterion.
Nevertheless, we will always assume in the sequel that the covariance data $\Sigmab$ are extendable and are known exactly.
We are interested in finding a line spectrum solution to \eqref{moment_eqns}, that is, a spectral measure of the form
\begin{equation}\label{sum_Dirac}
\d M(\thetab) = \sum_{\ell=1}^{L} Q_\ell \, \delta(\thetab-\thetab_\ell) \d\thetab
\end{equation}
where $\delta(\thetab-\thetab_\ell)$ is the Dirac delta with unit mass at $\thetab_\ell\in\Tbb^d$, and each $Q_\ell$ is a Hermitian positive semidefinite matrix, sometimes called \emph{densities} of $M$ \citep{kimsey2013truncated}.


\section{An existence result}\label{sec:existence}

In this section, we prove that a solution to \eqref{moment_eqns} of the form \eqref{sum_Dirac} always exists under the extendability assumption. The idea is drawn from \cite{lang1983spectral} which is built upon Carath\'{e}odory's theorem for convex hulls. However, the technique in the aforementioned paper is developed for scalar measures $(m=1)$, and cannot be directly applied to our matricial problem. For this reason, we shall first ``extend'' the measure $\d M$ in question to a product space and construct a scalar measure.

First, let us introduce the compact set $S:=\{ u\in\Cbb^m :\, \|u\|=1 \}$ where the notation $\|\cdot\|$ stands for the Euclidean $2$-norm if not otherwise specified. The set $S$ can also be viewed as the unit hypersphere in $\Rbb^{2m}$.
We need the following two lemmas in order to state the main result of this section.

\begin{lemma}\label{lem_prod_meas}
	Given a nonnegative matricial measure $\d M(\thetab)$ on $\Tbb^d$, there exists a nonnegative scalar measure $\d\nu(\thetab,u)$ on the product space $\Tbb^d \times S$ such that
	\begin{equation}\label{prod_meas}
	\d M(\thetab) = \int_{u\in S} uu^* \d\nu(\thetab,u).
	\end{equation}
\end{lemma}

\begin{pf}
	Let $\lambda:=\sum_{j,k=1}^{m}|m_{jk}|$ be a nonnegative scalar measure on $\Tbb^d$ where $m_{jk}$ (with a slight abuse of notation) is the $(j,k)$ element of the given matricial set function $M$ and $|\cdot|$ denotes the total variation measure. Then by the Radon-Nikodym theorem \citep{rudin1987real}, there exists a Hermitian matrix-valued measurable function $M'_\lambda$ on $\Tbb^d$ which is positive semidefinite $\lambda$-a.e. such that $\d M(\thetab)=M'_\lambda(\thetab) \d\lambda$.\footnote{In fact, in order for the density $M'_\lambda$ to exist, each $m_{jk}$ must be absolutely continuous with respect to the nonnegative scalar measure $\lambda$, and such $\lambda$ is obviously not unique.} For each $\thetab\in\Tbb^d$, do eigen-decomposition $M'_\lambda(\thetab)=\sum_{k=1}^{m}\varphi_k(\thetab)u_k(\thetab)u_k(\thetab)^*$ where $\{\varphi_k\}$ are nonnegative eigenvalues and $\{u_k\}$ are the orthonormal eigenvectors. The eigenvalues are measurable functions of $\thetab$ since they are the roots of the characteristic polynomial of $M'_\lambda(\thetab)$ whose entries are measurable, and the eigenvectors can also be chosen in a measurable way. Next, define the scalar measure
	$\d\nu(\thetab,u):=\sum_{k=1}^m \varphi_k(\thetab)\delta(u-u_k(\thetab))\d\lambda(\thetab)\d u$,
	and \eqref{prod_meas} can be readily verified:
	\begin{equation*}
	\begin{split}
	 & \int_{u\in S} uu^* \d\nu(\thetab,u) \\
	= & \sum_{k=1}^m \varphi_k(\thetab) \d\lambda(\thetab) \int_{u\in S} uu^* \delta(u-u_k(\thetab)) \d u \\
	= & \sum_{k=1}^{m}\varphi_k(\thetab)u_k(\thetab)u_k(\thetab)^* \d\lambda(\thetab) \\
	= & M'_\lambda(\thetab) \d\lambda(\thetab) = \d M(\thetab). 
	\end{split}	
	\end{equation*}
\end{pf}

\begin{remark}
	The scalar measure $\nu$ in \eqref{prod_meas} is in general not unique. For example, consider the univariate case $(m=1)$ where the set $S$ can be identified as $\Tbb$. Then \eqref{prod_meas} reduces to $\d m(\thetab) = \int_{u\in \Tbb} \d\nu(\thetab,u)$, and $\nu$ can be any product measure $m\times\mu$ such that $\mu$ has total mass one on $\Tbb$. 
\end{remark}

\begin{lemma}\label{lem_normalize}
	Assume that \eqref{prod_meas} holds for a nonnegative matricial measure $\d M$ and some nonnegative scalar measure $\d\nu$. If $\trace M(\Tbb^d)=1$, then the corresponding $\nu$ is a probability measure.
\end{lemma}
\begin{pf}
	We only need to show that $\nu$ has total mass $1$, and this can be done via straightforward computation:
	\begin{equation}
	\begin{split}
	1 = \trace M(\Tbb^d) & = \trace \int_{\Tbb^d} \d M(\thetab) \\
	 & = \int_{\Tbb^d\times S} \trace(uu^*) \d\nu(\thetab,u) \\
	 & = \int_{\Tbb^d\times S} \d\nu(\thetab,u),
	\end{split}	
	\end{equation}
	where the last equality holds because $\trace(uu^*)=\|u\|^2=1$ by the definition of the set $S$.
\end{pf}

Let $|\Lambda|$ denote the cardinality of the index set $\Lambda$. For a fixed $\thetab\in\Tbb^d$, stack the complex numbers $\{e^{i\innerprod{\kb}{\thetab}} :\, \kb\in\Lambda\}$ into a column vector $\ab(\thetab)\in\Cbb^{|\Lambda|}$ according to the lexicographical ordering. The elements of $\ab(\thetab)$ can be interpreted as the covariances of a scalar field having a line spectrum at $\thetab$ of unit mass.

\begin{theorem}[Line Spectrum Representation]\label{thm_represent}
	Given a finite set of extendable covariances $\{\Sigma_{\kb}\}_{\kb\in\Lambda}$ such that $\trace \Sigma_\zerob=1$, there exists a solution to \eqref{moment_eqns} of the form
	\begin{equation}\label{represent_Dirac}
	\d M(\thetab) = \sum_{\ell=1}^{L_1} r_\ell u_\ell u_\ell^* \, \delta(\thetab-\thetab_\ell) \d\thetab
	\end{equation}
	where the integer $L_1=m^2|\Lambda|$, each $r_\ell$ is a nonnegative real number, $u_\ell\in S$, $\thetab_\ell\in\Tbb^d$, and $\sum_{\ell=1}^{L_1} r_\ell=1$.
\end{theorem}

\begin{pf}
	Since the covariances $\{\Sigma_{\kb}\}$ are extendable by assumption, there exists a nonnegative matricial measure $\d M_0$ such that \eqref{moment_eqns} holds. Appealing to Lemma \ref{lem_prod_meas}, we can rewrite the moment equations as
	\begin{equation}\label{moment_eqns_modified}
	\int_{\Tbb^d\times S} e^{i\innerprod{\kb}{\thetab}} uu^* \d\nu_0(\thetab,u) = \Sigma_{\kb} \text{ for all } \kb\in\Lambda
	\end{equation}
	for some nonnegative scalar measure $\d\nu_0$.
	Using the vector notation, \eqref{moment_eqns_modified} can be put in a compact form
	\begin{equation}\label{moment_eqns_vectorized}
	\int_{\Tbb^d\times S} (\ab(\thetab)\otimes I_m) uu^* \d\nu_0(\thetab,u) = \Sigmab
	\end{equation}
	where the matrix $\Sigmab$ contains the covariances $\{\Sigma_{\kb}\}$ in accordance with the ordering in $\ab(\thetab)$. Define the function $\Ab(\thetab,u):=(\ab(\thetab)\otimes I_m)uu^*$ to ease the notation. Notice that $\Sigma_\zerob=M_0(\Tbb^d)$ by definition. Therefore, the condition of unit trace makes Lemma \ref{lem_normalize} applicable, and we can conclude that $\nu_0$ is a probability measure, which combined with the relation \eqref{moment_eqns_vectorized}, implies that $\Sigmab$ is in the convex hull\footnote{In fact, due to the integral, $\Sigmab$ belongs to the closure of $\conv(A)$. However, since $A$ is easily seen to be compact and the convex hull of a compact set (in a finite-dimensional space) is again compact, $\conv(A)$ and its closure coincide.} of the set
	\begin{equation}\label{set_A}
	A:=\{\Ab(\thetab,u) :\, (\thetab,u)\in\Tbb^d\times S\}.
	\end{equation}
	Since $\Sigmab$ lives in a vector space of real dimension\footnote{Take into account the symmetry $\Sigma_{-\kb}=\Sigma_{\kb}^*$, the fact that $\Sigma_\zerob$ is Hermitian, and the affine condition $\trace\Sigma_\zerob=1$ which further reduces the dimension by $1$.} $L_1-1=m^2|\Lambda|-1$, by Carath\'{e}odory's theorem for convex hulls, $\Sigmab$ can be written as the convex combination of at most $L_1$ points in $A$, that is,
	\begin{equation}\label{moment_eqns_discrete}
	\Sigmab = \sum_{\ell=1}^{L_1} r_\ell \Ab(\thetab_\ell,u_\ell)
	\end{equation}
	which corresponds to the measure
	\begin{equation}
	\d\nu = \sum_{\ell=1}^{L_1} r_\ell \,\delta(\thetab-\thetab_\ell) \,\delta(u-u_\ell) \d\thetab \d u.
	\end{equation}
	Finally, the measure $\d M$ in \eqref{represent_Dirac} can be recovered from $\d\nu$ via the relation \eqref{prod_meas}.
\end{pf}

\begin{remark}
	The representation \eqref{sum_Dirac} can be obtained from \eqref{represent_Dirac} by grouping the summands with the same $\thetab_\ell$.
	In addition, the normalization condition $\trace \Sigma_\zerob=1$ in Theorem $\ref{thm_represent}$ is of no restriction because whenever the underlying random field is not trivial (that is, $\Sigma_\zerob$ is not equal to the all-zero matrix), the covariances can be rescaled by a factor of $1/(\trace \Sigma_\zerob)$. The only difference is that the coefficients $r_\ell$ will also be rescaled so that they sum to $\trace \Sigma_\zerob$.
\end{remark}

\begin{remark}
	The paper \cite{kimsey2013truncated} also considers \emph{finitely-atomic} solutions to matrix-valued moment problems on multidimensional domains, and it gives the existence of a \emph{minimum-rank} solution but under much more sophisticated conditions.
	Our result here is weaker, but the mathematics leads to it is conceptually simpler and it will be useful in our later developments.
\end{remark}

\section{Uniqueness of the representation}\label{sec:uniqueness}

It is well known that in the scalar unidimensional case $(m=d=1)$, when the covariance sequence $\sigmab=\{\sigma_0,\dots,\sigma_n\}$ lies in the interior of the dual cone $\Cfrak_+$, the line spectrum representation is never unique. Notice that the interior condition for the covariance sequence here is very simple, as it amounts to the positive definiteness of the Toeplitz matrix
\begin{equation}\label{Toeplitz_mat}
T(\sigmab):=\bmat \sigma_0&\sigma_1^*&\sigma_2^*&\cdots&\sigma_n^* \\
\sigma_1&\sigma_0&\sigma_1^*&\cdots&\sigma_{n-1}^* \\
\sigma_2&\sigma_1&\sigma_0&\cdots&\sigma_{n-2}^* \\
\vdots&\vdots&\ddots&\ddots&\vdots \\
\sigma_n&\sigma_{n-1}&\cdots&\sigma_1&\sigma_0\emat.
\end{equation}
Uniqueness of the line spectrum representation holds true when the matrix $T(\sigmab)$ is singular, and this fact is called Carath{\'e}odory--Fej{\'e}r--Pisarenko decomposition in \citep{georgiou2007caratheodory}, also termed Vandermonde decomposition in signal processing literature \citep[see e.g.,][]{yang2018frequency}. Following this direction, one seeks to decompose the matricial spectrum in question into the form of ``signal plus noise'':
\begin{equation}\label{CFP_decomp}
\d M(\thetab) = \d \tilde{M}(\thetab) + \alpha R\, \d\thetab,
\end{equation}
where $\d\tilde{M}$ stands for the sinusoidal signal that corresponds to the covariance data on the boundary of the dual cone $\partial \Cfrak_+$, $R\geq0$ is a constant variance matrix attributed to an i.i.d. noise process, and $\alpha$ is a nonnegative real number. In the multivariate setting $(m>1)$, there is a degree of freedom to choose the $R$ matrix, and once it is fixed, one can always push the covariance data $\Sigmab$ to the boundary of the dual cone by the subtraction $\tilde{\Sigma}_\zerob=\Sigma_\zerob-\alpha R$ for a suitable $\alpha\geq0$. Notice that in order to qualify the decomposition \eqref{CFP_decomp} as ``canonical'', we need to pose the uniqueness question of the line spectrum representation for \emph{any} covariance data $\Sigmab\in\partial\Cfrak_+$, as will be discussed next.

By the definition of the dual cone \eqref{dual_cone}, $\Sigmab\in\partial\Cfrak_+$ means that there exists some nonzero $P\in\Pfrak_+$ such that $\innerprod{\Sigmab}{\Pb}=0$.
Using the general spectral representation \eqref{Sigma_spec}, we can rewrite the inner product \eqref{inner_prod} as
\begin{equation}\label{inner_prod_integral}
\begin{split}
\innerprod{\Sigmab}{\Pb} & = \trace \sum_{\kb\in\Lambda} P_\kb^* \int_{\Tbb^d} e^{i\innerprod{\kb}{\thetab}} \d M \\
 & = \trace \int_{\Tbb^d} P(e^{i\thetab}) \d M \\
 & = \int_{\Tbb^d} \trace[ P(e^{i\thetab}) M'_\lambda(\thetab) ] \d\lambda.
\end{split}
\end{equation}
Recall also the basic fact that for two positive semidefinite matrices $A,B$, $\trace(AB)=0$ if and only if $AB=0$. Therefore, $\innerprod{\Sigmab}{\Pb}=0$ if and only if $P(e^{i\thetab}) M'_\lambda(\thetab)=\zerob$ $\lambda$-a.e. It means that with possible exceptions on a $\lambda$-null set, each column of the matrix $M'_\lambda(\thetab)$ belongs to the kernel of $P(e^{i\thetab})$. In particular, whenever $P(e^{i\thetab})>0$, it must happen that $M'_\lambda(\thetab)=\zerob$. Hence the support of $\d M$ is contained in the zero set
\begin{equation}
\Zcal(P):=\{ \thetab\in\Tbb^d \,:\, \det P(e^{i\thetab})=0 \}.
\end{equation}
The next proposition gives a sufficient condition for the uniqueness of the line spectrum representation.

\begin{proposition}\label{prop_unique_sufficient}
	For $\Sigmab\in\partial\Cfrak_+$ such that $\innerprod{\Sigmab}{\Pb}=0$ for some $P\in\Pfrak_+$, if any finite collection of vectors from the set
	\begin{equation}\label{corr_vec_zero_set}
	\{ \ab(\thetab) \,:\, \thetab\in\Zcal(P) \}
	\end{equation}
	are linearly independent, then the line spectrum representation \eqref{sum_Dirac} for $\Sigmab$ is unique.
\end{proposition}


\begin{pf}
	The proof uses elementary techniques from linear algebra.
	By Theorem \ref{thm_represent}, the covariances $\Sigmab$ admits a line spectrum representation \eqref{represent_Dirac}, which can be written in a matrix form as
	\begin{equation}
	\begin{split}
	\Sigmab & = \sum_{\ell=1}^L \left( \ab(\thetab_\ell) \otimes I_m \right) Q_\ell \\
	 & = \left\{ \bmat \ab(\thetab_1) & \cdots & \ab(\thetab_L) \emat \otimes I_m \right\} \times \bmat Q_1 \\ \vdots \\ Q_L \emat.
	\end{split}
	\end{equation}
	The ``coefficient matrix'' on the left still has linearly independent columns due to a property of the Kronecker product. If $\Sigmab$ has another representation of this form, then the density matrices $Q_\ell$ corresponding to the common set of $\thetab_\ell$ must be identical while the rest $Q_\ell$ must be all zero as a consequence of linear independence, meaning that the above representation is unique.
\end{pf}

\begin{remark}
	In the scalar case, the condition in Proposition~$\ref{prop_unique_sufficient}$ is also necessary for the uniqueness of the line spectrum representation \citep[cf.][]{lang1983spectral}, as one can show that if a finite subset of vectors from \eqref{corr_vec_zero_set} are linearly dependent, then there exists some $\sigmab\in\partial\Cfrak_+$ having two different representations.	
	Due to the additional structure in our matricial problem, this point seems nontrivial, as partially illustrated in the following example.
\end{remark}

\begin{example}\label{ex_degenerate}
	Given $\Sigmab\in\partial\Cfrak_+$, combining the representation \eqref{represent_Dirac} with the relation \eqref{inner_prod_integral}, we know that the spectral lines must be located in the zero set of the determinant of some $P\in\Pfrak_+$, and that $P(e^{i\thetab_\ell})Q_\ell=\zerob$. 
	
	Now, let us consider a degenerate case in which $P(e^{i\thetab})\equiv P_0$ where $P_0\geq0$ is singular. Obviously, we have $\Zcal(P)=\Tbb^d$, and there exists a finite set of linearly dependent vectors $\{\ab(\theta_\ell)\}$. Following \citet[Appendix C]{lang1983spectral}, we have two different representations for the same covariance data
	\begin{equation}
	\Sigmab = \sum_{b_\ell>0} \ab(\thetab_\ell) \otimes (b_\ell u_0 u_0^*) = \sum_{b_\ell<0} \ab(\thetab_\ell) \otimes (-b_\ell u_0 u_0^*)
	\end{equation}
	where $\{b_\ell\}$ are the nonzero coefficients for the linear combination of $\{\ab(\thetab_\ell)\}$, and $u_0\in S$ such that $P_0 u_0=0$.

	The above construction uses essentially scalar thinking. When $\det P(e^{i\thetab})$ is not constantly zero, this type of construction seems nontrivial and is not known to the author.
\end{example}

In order to exclude such degenerate cases as in Example~\ref{ex_degenerate}, we introduce the following definition.
\begin{definition}
	A covariance multisequence $\Sigmab\in\partial\Cfrak_+$ is called nondegenerate if there exists $P\in\Pfrak_+$ whose determinant is not identically zero on $\Tbb^d$ such that $\innerprod{\Sigmab}{\Pb}=0$.
\end{definition}
From the condition in Proposition~\ref{prop_unique_sufficient}, it follows that in order for the line spectrum representation of any nondegenerate $\Sigmab\in\partial\Cfrak_+$ to be unique, the zero set $\Zcal(P)$ cannot contain more than $|\Lambda|$ elements for any $P\in\Pfrak_+$ such that $\det P(e^{i\thetab})\not\equiv0$.
As discussed in \cite{lang1983spectral} for the scalar case, such a uniqueness condition is very strong and does not hold in general when $d>1$.

\subsection{Specialization to the $1$-d case}

In this subsection, we consider the special $1$-d case and provide a new uniqueness result of the line spectrum representation in the bivariate setting $(m=2)$ as well as a computational procedure to obtain such representation. These can be seen as a supplement to \citet[Section VI]{georgiou2007caratheodory}. Notice here that the index set $\Lambda=\{-n,\dots,0,\dots,n\}$, and $|\Lambda|=2n+1$.

\begin{corollary}\label{cor_1d_unique}
	Fix $d=1$. The line spectrum representation is unique for any nondegenerate $\Sigmab=(\Sigma_0,\dots,\Sigma_n)\in\partial\Cfrak_+$ if $m=1,2$ or $m=3$ and $n\leq1$.
\end{corollary}

\begin{pf}
	By nondegeneracy, there is a $P\in\Pfrak_+$ such that $\det P(e^{i\theta})\not\equiv0$ and $\innerprod{\Sigmab}{\Pb}=0$. When $d=1$, the nonnegative matricial polynomial $P$ admits a spectral factorization $P(z)=A(z)A^*(z)$ where $A(z)=\sum_{k=0}^n A_k z^{-k}$ for $A_k\in\Cbb^{m\times m}$. It follows that $\det P(z)=\det A(z) [\det A(z)]^*$ is a nonnegative Laurent polynomial of degree no more than $mn$ and it can only have $mn$ or fewer roots on the unit circle.	In other words, the zero set $\Zcal(P)$ has a finite number of elements $\theta_1,\dots,\theta_L$ with $L\leq mn$. Appealing to Proposition~\ref{prop_unique_sufficient}, the line spectrum representation is unique if the $(2n+1) \times L$ Vandermonde matrix
	\begin{equation}
	\bmat \ab(\theta_1) & \cdots & \ab(\theta_L) \emat
	\end{equation}
	has linearly independent columns, which holds true if and only if $L\leq 2n+1$. Therefore, a sufficient condition is $mn\leq 2n+1$ which is satisfied if $m=1,2$ or $m=3$ and $n\leq1$.	
\end{pf}

The above corollary can be reformulated in terms of the Vandermonde decomposition of positive semidefinite block-Toeplitz matrices, and in particular, we are interested in the case with $2\times2$ blocks. More precisely, given the covariance data $\Sigmab=(\Sigma_0,\dots,\Sigma_n)$, form the block-Toeplitz matrix
\begin{equation}\label{moment_eqns_matrix}
T(\Sigmab) = \bmat\Sigma_0&\Sigma_1^*&\Sigma_2^*&\cdots&\Sigma_n^* \\
\Sigma_1&\Sigma_0&\Sigma_1^*&\cdots&\Sigma_{n-1}^* \\
\Sigma_2&\Sigma_1&\Sigma_0&\cdots&\Sigma_{n-2}^* \\
\vdots&\vdots&\ddots&\ddots&\vdots \\
\Sigma_n&\Sigma_{n-1}&\cdots&\Sigma_1&\Sigma_0\emat.
\end{equation}
Notice that $\Sigmab\in\partial\Cfrak_+$ if and only if $T(\Sigmab)$ is positive semidefinite and singular.
Then one can rewrite the $1$-d moment equations in a compact form
\begin{equation}\label{G_filter}
T(\Sigmab) = \int_\Tbb G(\theta) \d M G^*(\theta)
\end{equation}
where $G(\theta) := \gb(\theta) \otimes I_2$ and
\begin{equation}\label{g_vec}
\gb(\theta) := [\, 1, e^{i\theta}, \dots, e^{in\theta} \,]^\top\in\Cbb^{n+1}.
\end{equation}
Substitute the solution form \eqref{sum_Dirac} into \eqref{moment_eqns_matrix}, we obtain the decomposition of the block-Toeplitz matrix
\begin{equation}\label{Vandermonde_decomp_block_Toeplitz}
\begin{split}
T(\Sigmab) & = \sum_{\ell=1}^L G(\theta_\ell) Q_\ell G^*(\theta_\ell) \\
 & = \bmat G(\theta_1) & \cdots & G(\theta_L) \emat
 \bmat Q_1&&\\&\ddots&\\&&Q_L\emat
 \bmat G^*(\theta_1) \\ \vdots \\ G^*(\theta_L) \emat
\end{split}
\end{equation}
where $L\leq2n$, $Q_\ell\geq0$, and $\theta_\ell\in\Tbb$ $(\ell=1,\dots,L)$ are distinct. Moreover,  by Corollary~\ref{cor_1d_unique}, the decomposition is unique if the covariance sequence $\Sigmab\in\partial\Cfrak_+$ is nondegenerate.

Next, we describe how to compute the Vandermonde decomposition via linear algebraic techniques. The following computational procedure has been given in \cite{gurvits2002largest}, but we include it here for the sake of completeness.
Notice that the procedure works for any positive integer $m$ (size of the blocks) even if the uniqueness of the decomposition may fail to be true. Assume that $T(\Sigmab)$ is positive semidefinite having rank $r<m(n+1)$. Then it admits a rank factorization $T(\Sigmab)=VV^*$ where
\begin{equation}
V=\bmat V_0 \\ \vdots \\ V_n\emat \in \Cbb^{m(n+1)\times r}
\end{equation}
and each block $V_k$ is of size $m\times r$. Let $V_{-0}$ and $V_{-n}$ denote the $mn\times r$ matrices obtained from $V$ by removing the first and the last block row, respectively. Due to the block-Toeplitz structure, we have $V_{-0}V_{-0}^* = T(\Sigma_0,\dots,\Sigma_{n-1}) = V_{-n}V_{-n}^*$. Thus by \citet[Theorem 7.3.11]{horn2013matrix}, there exists a $r\times r$ unitary matrix $U$ such that
\begin{equation}\label{relation_U}
V_{-0}=V_{-n}U.
\end{equation}
It follows that the blocks of $V$ satisfies $V_k = V_0 U^k$ $(k=1,\dots,n)$, and the covariance data can be expressed as $\Sigma_k = V_k V_0^* = V_0 U^k V_0^*$. Introduce the eigen-decomposition
\begin{equation}\label{eigen_decomp_U}
U = \tilde{U} \diag\{e^{i\theta_1},\dots,e^{i\theta_r}\} \tilde{U}^*
\end{equation}
where $\tilde{U}$ is also unitary and $\{e^{i\theta_\ell}\}$ are eigenvalues of unit modulus. Then we can further write
\begin{equation}\label{decomp_Sigma_k}
\begin{split}
\Sigma_k & = V_0 \tilde{U} \diag\{e^{ik\theta_1},\dots,e^{ik\theta_r}\} \tilde{U}^* V_0^* \\
 & = \sum_{\ell=1}^r e^{ik\theta_\ell} (V_0\tilde{U})_{:,\ell}\, (V_0\tilde{U})_{:,\ell}^*
\end{split}
\end{equation}
where $A_{:,\ell}$ is the Matlab notation standing for the $\ell$-th column of a matrix $A$. A distinct feature in the multivariate case is that $U$ may have eigenvalues of multiplicity larger than $1$. After combining terms corresponding to identical eigenvalues, we arrive at the desired decomposition \eqref{Vandermonde_decomp_block_Toeplitz} where
\begin{equation}
\Sigma_k = \sum_{\ell=1}^L e^{ik\theta_\ell} Q_\ell.
\end{equation}
Hence in general, $L\leq r$.

From the above discussion, it is clear that important parameters $\{\theta_\ell,\,Q_\ell\}$ of the Vandermonde decomposition of $T(\Sigmab)$ are encoded in the eigen-decomposition of the unitary matrix $U$. In principle, such $U$ can be constructed explicitly \citep{horn2013matrix}. However, a more efficient approach is the following. Multiplying $V_{-n}^*$ from the left to both sides of \eqref{relation_U} and using the eigen-decomposition \eqref{eigen_decomp_U}, we can obtain the relation
\begin{equation}
V_{-n}^*V_{-0} \tilde{U}_{:,\ell} = e^{i\theta_\ell} V_{-n}^*V_{-n} \tilde{U}_{:,\ell}.
\end{equation}
Therefore, the unitary matrix $\tilde{U}$ and the eigenvalues $\{e^{i\theta_\ell}\}$ can be obtained by solving the generalized eigenvalue problem of the ordered matrix pair $(V_{-n}^*V_{-0},\,V_{-n}^*V_{-n})$.


\section{Application to frequency estimation}\label{sec:freq_est}


Consider $\Cbb^2$ vectorial measurements $y$ obeying the model
\begin{equation}\label{signal_model}
y(t) = \sum_{\ell=1}^L s_\ell \, e^{i\theta_\ell t} + w(t)
\end{equation}
where, $t=0,1,\dots,n$, $L$ is the number of sources, $\{s_\ell\}$ are $2$-d complex vectorial amplitudes, $\{\theta_\ell\in\Tbb\}$ are unknown (but fixed) frequencies, $x(t):=\sum_{\ell=1}^L s_\ell \, e^{i\theta_\ell t}$ is the signal component, and $w(t)$ is a noise process.
The measurement equation can be put in a matrix form
\begin{equation}\label{y_meas_vec}
\yb := \bmat y(0) \\ \vdots \\ y(n) \emat = \underbrace{\bmat G(\theta_1) & \cdots & G(\theta_L) \emat \bmat s_1 \\ \vdots \\ s_L \emat}_{\xb} + \bmat w(0) \\ \vdots \\ w(n) \emat
\end{equation}
where the block columns $\{G(\theta_\ell)\}$ are defined after \eqref{G_filter}.
We first consider the noiseless case where we have $\yb=\xb$ exactly.

A popular approach nowadays in frequency estimation involves the notion of the atomic norm of the signal $\xb$ \citep[see, e.g.,][]{yang2018sparse}. The elements in the set $\{G(\theta)\,:\,\theta\in\Tbb\}$ are called ``atoms'' which can be viewed as \emph{over-complete} basis functions with a continuous parameter $\theta\in\Tbb$. The atomic norm of $\xb$ is defined as
\begin{equation}\label{def_AN_xb}
\begin{split}
\|\xb\|_\Acal := \inf_{s_\ell,\theta_\ell} \left\{ \sum_{\ell} \|s_\ell\| :\, \xb = \sum_{\ell} G(\theta_\ell) s_\ell, \right. \\ \left. \theta_\ell\in\Tbb,\ s_\ell\neq 0\in \Cbb^2 \right\}
\end{split}
\end{equation}
which is the continuous counterpart of the $\ell^1$ norm, so that it promotes sparsity in the sense that the decomposition of the signal $\xb$ contains as few terms as possible.

The definition of the atomic norm does not indicate a way to compute it. The next result addresses this issue and shows that the atomic norm admits a semidefinite programming (SDP) formulation (modulo a rank condition) and can be computed efficiently using standard tools for convex optimization \citep{bv_cvxbook}. The proof is deferred to the appendix.

\begin{theorem}\label{thm_AN_noiseless}
	Given the noiseless measurements $\xb$ of the complex sinusoids, let $p$ be the optimal value of the semidefinite programming
	\begin{subequations}\label{AN_semidef_program}
	\begin{align}
	& \underset{b,\Sigmab}{\text{minimize}}
	& & \frac{1}{2}b + \frac{1}{2} \trace\Sigma_0 \label{obj_noiseless} \\
	& \text{subject to}
	& & \bmat b&\xb^*\\\xb&T(\Sigmab) \emat \geq 0. \label{LMI_constraint}
	\end{align}
	\end{subequations}
	Then the atomic norm $\|\xb\|_\Acal\geq p$. Moreover, if the minimizer $(\hat{b},\hat{\Sigmab})$ of \eqref{AN_semidef_program} is such that $\hat{r}:=\rank T(\hat{\Sigmab})\leq n+1$ (half of the size of $T(\hat{\Sigmab})$) with $\hat{\Sigmab}$ nondegenerate, then $\|\xb\|_\Acal=p$.
\end{theorem}

The above theorem suggests a way of doing frequency estimation (in the noiseless case) by first solving the optimization problem \eqref{AN_semidef_program}. Then given the optimal $T(\hat{\Sigmab})$, the unknown frequencies $\{\theta_\ell\}$ can be recovered via computing its Vandermonde decomposition.

\begin{remark}\label{rem:interp_cov_mat}
	The matrix $T(\Sigmab)$ can be interpreted as the signal covariance matrix $\E(\xb\xb^*)$ if the amplitudes $\{s_\ell\}$ are modeled as zero-mean random vectors such that $\E(s_\ell s_\ell^*)=Q_\ell\geq0$ and $\E(s_\ell s_k^*)=0$ if $\ell\neq k$. Then one can easily see from \eqref{y_meas_vec} that $\E(\xb\xb^*)$ admits the Vandermonde decomposition \eqref{Vandermonde_decomp_block_Toeplitz}. However, the optimization approach involving the atomic norm differs from traditional subspace methods which directly estimate the covariance matrix from the measurements, in that the block-Toeplitz structure of the covariance matrix is explicitly enforced and a low-rank solution is sought.
\end{remark}

\begin{remark}
	Since no statistical assumptions have been made on the signal model \eqref{signal_model}, the proposed method for frequency estimation can be labeled as ``deterministic''. In this context, an alternative formulation given the vectorial measurements is the following. Collect the noiseless measurements of channel $k$ $(=1,2)$ into a column vector $\xb_k:=[x_k(0),\dots,x_k(n)]^\top\in\Cbb^{n+1}$, called a \emph{snapshot}. One can then pose the frequency estimation problem given two (in general, multiple) snapshots. Similarly, the atomic norm of $\Xb:=[\xb_1,\xb_2]\in\Cbb^{(n+1)\times2}$ can be defined as
	\begin{equation}\label{atomic_norm_Xb}
	\begin{split}
	\|\Xb\|_\Acal := \inf_{s_\ell,\theta_\ell} \left\{ \sum_{\ell}\|s_\ell\| \,:\, \Xb=\sum_{\ell} \gb(\theta_\ell) s_\ell^\top, \right. \\ \left. \theta_\ell\in\Tbb,\ s_\ell\neq 0\in \Cbb^2 \right\}
	\end{split}
	\end{equation}
	with $\gb(\theta)$ in \eqref{g_vec}. It is clear that the above definition is equivalent to \eqref{def_AN_xb} up to a rearrangement of the data. Moreover, $\|\Xb\|_\Acal$ also admits a SDP characterization similar to that in Theorem~\ref{thm_AN_noiseless} \citep[see][]{yang2016exact}. When the rank condition in Theorem~\ref{thm_AN_noiseless} is satisfied, the two formulations are equivalent and the result of performance guarantee in \citet[Theorem 4]{yang2016exact} can be applied to conclude \emph{exact frequency recovery} when the unknown frequencies are sufficiently separated. More precisely, the separation condition can be expressed as
	\begin{equation}\label{separat_cond}
	\frac{\Delta\theta}{2\pi} \geq \frac{4}{n}
	\end{equation}
	where $\Delta\theta := \min_{1\leq k\neq \ell\leq L} \min \{ |\theta_k-\theta_\ell|, 2\pi-|\theta_k-\theta_\ell|\}$ is the minimum circular distance between any two elements in the set $\{\theta_\ell\}_{\ell=1}^{L}$. The minimum frequency separation is also known as ``resolution'' in the literature of frequency estimation. The quantity $4/n$ in \eqref{separat_cond} is often larger than those of FFT-based methods and subspace methods, meaning that the atomic norm approach has a lower resolution, which is its major drawback. However, it also has great advantages such as admitting a convex formulation (hence solvable) and automatic detection of the number $L$ of sinusoids (via the Vandermonde decomposition of the optimal $T(\hat\Sigmab)$).
\end{remark}


Noiseless measurements are of course just theoretical idealization since in practice noise is ubiquitous. Thus, it is of great practical interest to investigate the frequency estimation problem in the noisy case, which is also known as \emph{atomic norm denoising} \citep{bhaskar2013atomic}. Typically, one uses the atomic norm as a regularization term and sets up the following optimization problem:
\begin{equation}\label{AN_regularized_LS}
\underset{\xb}{\text{minimize}} \ f(\xb) := \frac{1}{2}\|\xb-\yb\|^2 + \tau \|\xb\|_\Acal \\
\end{equation}
where the regularization parameter $\tau$ should be chosen properly. Given Theorem~\ref{thm_AN_noiseless} and assuming the rank condition in it, for each $\xb$ we can write
\begin{equation}
f(\xb) = \min_{b,\Sigmab} h(\xb,b,\Sigmab) \text{ s.t. } \eqref{LMI_constraint}
\end{equation}
where
\begin{equation}
h(\xb,b,\Sigmab) := \frac{1}{2}\|\xb-\yb\|^2 + \frac{\tau}{2} (b + \trace\Sigma_0).
\end{equation}
Therefore, the problem \eqref{AN_regularized_LS} is equivalent to the following SDP:
\begin{equation}\label{semidef_program_noisy}
\underset{\xb,b,\Sigmab}{\text{minimize}}\ h(\xb,b,\Sigmab) \quad \text{subject to}\ \eqref{LMI_constraint}.
\end{equation}
It now remains to choose the regularization parameter $\tau$. According to \citet{bhaskar2013atomic,li2015off}, under the assumption that the additive noise $w$ is zero-mean i.i.d.~Gaussian of variance $\sigma_w^2$, the choice
\begin{equation}\label{value_of_tau}
\tau = \sigma_w \sqrt{(n+1)\left[2+\log(n+1)+\sqrt{4\log(n+1)}\right]}
\end{equation}
leads to a stable recovery of the signal $\xb$. Notice however that in practice, the noise variance $\sigma_w^2$ is unknown and must be estimated from the measurements $y$. Such estimation can be carried out in the fashion of \citet[Section~V]{bhaskar2013atomic}. 
More precisely, for each (scalar) measurement channel $k=1,2$, compute the \emph{standard biased covariance estimates}
\begin{equation}\label{standard_cov_estimates}
\hat\sigma_k(j) = \frac{1}{n+1} \sum_{t=0}^{n-j} y_k(t+j)\, y_k(t)^*
\end{equation}
of lag $j=0,1,\dots$ up to $\tilde{n}\approx n/3$, and form the empirical covariance matrices $\hat{T}_k=T(\hat{\sigma}_k(0),\hat{\sigma}_k(1),\dots,\hat{\sigma}_k(\tilde{n}))$ where the notation conforms with \eqref{Toeplitz_mat}. Then perform a further average $\hat{T}=(\hat{T}_1+\hat{T}_2)/2$ and the noise variance is estimated via averaging the smallest $25\%$ of the eigenvalues of $\hat{T}$. The reason behind these operations is that under suitable assumptions \citep[see][Chapter~4]{stoica2005spectral}, the true covariance matrix of a sinusoidal signal in noise is of the form \eqref{additive_decomp_Toeplitz}, where $\alpha=\sigma_w^2$ and the singular summand has a rank much lower than its dimension, while the estimate $\hat{T}$ should be close to its theoretical value in norm.

Next, we present numerical examples of the frequency estimation problem approached via atomic norm minimization.

\subsection{Numerical simulations}



First, we present an example in the noiseless case.
The signal length is set as $n=64$ the number of hidden frequencies is $L=4$. The true frequency vector is randomly generated as $\thetab=[-0.3419,-0.0643,0.9193,1.3155]\in\Tbb^4$. Both the real and imaginary parts of each $2$-d complex amplitude are obtained via the $\verb|rand(2,1)|$ function in Matlab. Given these parameters, the measurements can be generated according to \eqref{signal_model} (with $w(t)\equiv0$).
Then we proceed to solve the SDP \eqref{AN_semidef_program} using CVX, a package for specifying and solving convex programs \citep{gb08,cvx} which in turn calls SDPT3 \citep{toh1999sdpt3}. The optimal value of the objective is $8.8901$, and the optimal $T(\hat{\Sigmab})$ has numerical rank $4$ where eigenvalues no greater than $\varepsilon=10^{-4}$ are treated as numerically zero. In particular, the fifth largest eigenvalue here is $7.5640\times10^{-8}$. The frequency estimate $\hat{\thetab}$ is computed from the Vandermonde decomposition of $T(\hat{\Sigmab})$ described in the previous section. The absolute error $\|\hat{\thetab}-\thetab\|$ of the estimate is $3.1644\times10^{-11}$ (essentially zero), meaning that the frequency recovery is exact.

Next, we consider the noisy case. The signal length $n$ is still fixed to $64$, and we do experiments under different choices of the number of sinusoidal components $L$ and the signal-to-noise ratio ($\SNR$). The latter is defined as $20\log_{10}(\sigma/\sigma_w)$ dB where the signal standard deviation is $\sigma=\sqrt{1/6}$ which comes from twice the variance of the uniform distribution $U[0,1]$. 
Once the parameters $L$ and $\SNR$ are chosen, we can generate the amplitude vectors, the frequencies, and the additive complex Gaussian noise to produce the measurements $y$. Then we run the procedure described around \eqref{standard_cov_estimates} to estimate the noise variance, which can be used to compute the regularization parameter via \eqref{value_of_tau}. We are now ready to solve \eqref{semidef_program_noisy} using CVX and the frequency estimate is obtained via the Vandermonde decomposition of the optimal $T(\hat{\Sigmab})$.

\begin{figure}[!h]
	\begin{centering}
		\includegraphics[width=0.5\textwidth]{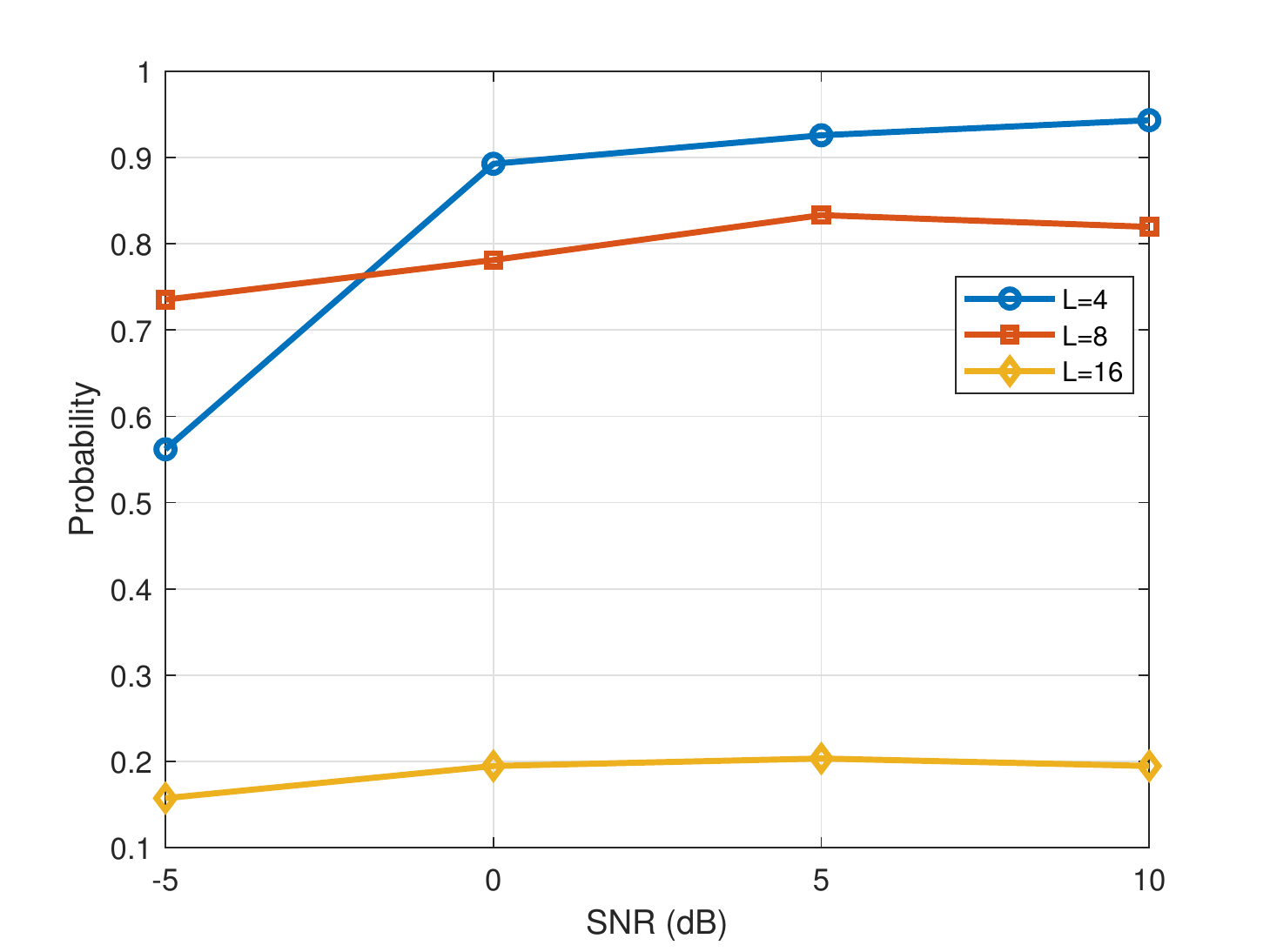}
		\caption{Probabilities of correctly recovering the number of hidden frequencies $L$ under different SNRs while the signal length $n=64$ is fixed.}
		\label{fig:p_recov_vs_SNR}
	\end{centering}
\end{figure}

In view of Remark~\ref{rem:rank} in the appendix, the rank of $T(\hat{\Sigmab})$ is expected to equal to $L$, the true number of unknown frequencies.
It is noted however, that in the noisy case, the optimal $T(\hat{\Sigmab})$ may have a ``wrong'' rank which corresponds to either missing or spurious frequency estimates. In order to investigate how often this happens, under each parameter configuration ($L$ and the $\SNR$), we do a Monte-Carlo simulation of a number of trials until $50$ correct rank recoveries are achieved. The we define the ``probability'' of such correct recovery as $50/\verb|tot_iter|$ where the variable \verb|tot_iter| denotes the number of total trials in one Monte-Carlo simulation which may vary as the parameter configuration changes. The simulation results in this respect are reported in Fig.~\ref{fig:p_recov_vs_SNR}. A general trend is that the correct rank is recovered more and more often as the SNR improves. The figure also shows that the method basically breaks down in the case of $L=16$ as the probability of correctly recovering the rank is around only $20\%$, which is very likely due the the violation of the separation condition \eqref{separat_cond}.

\begin{figure}[!h]
	\begin{centering}
		\includegraphics[width=0.5\textwidth]{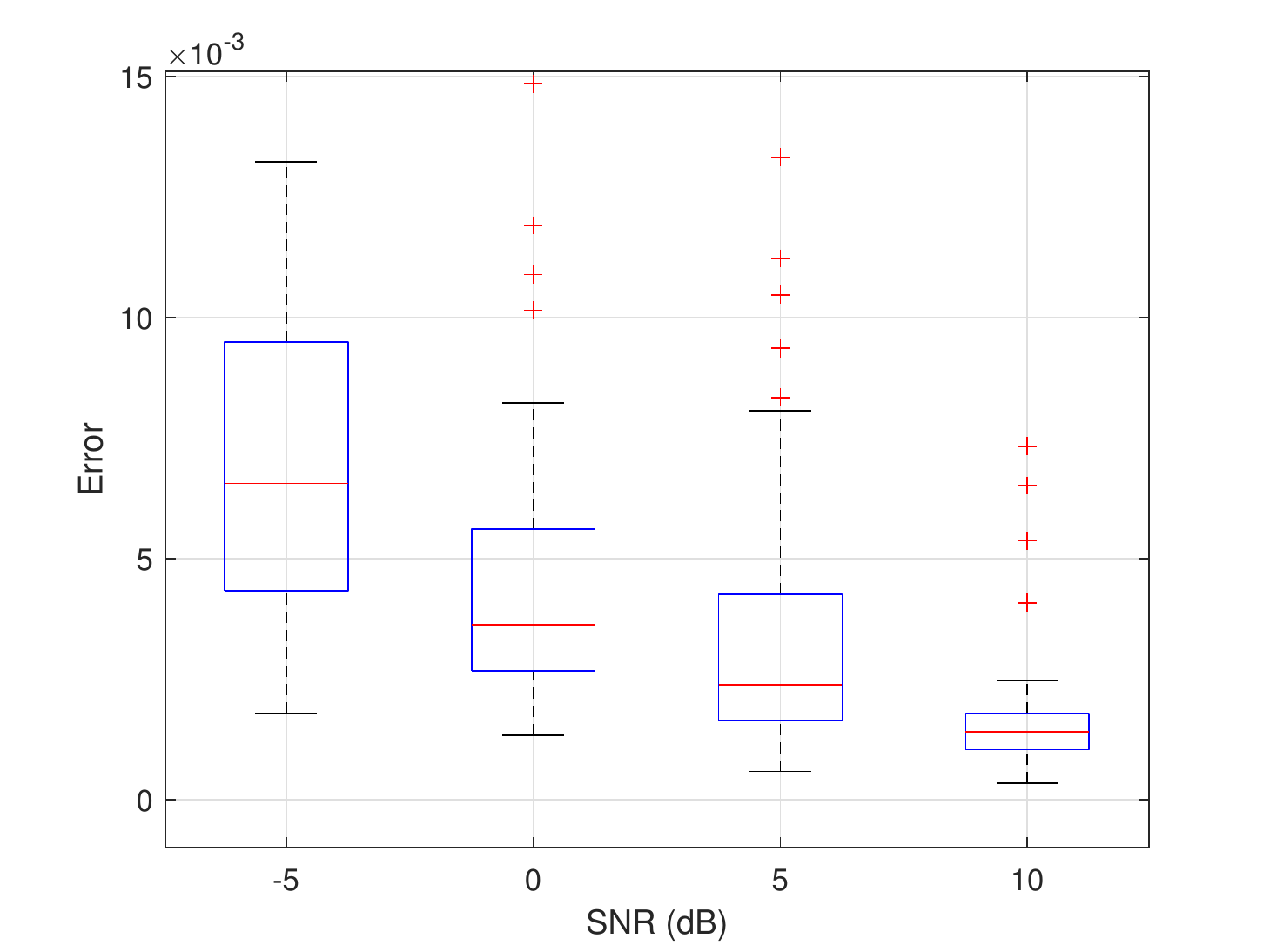}
		\caption{Errors of frequency estimation versus the SNR while the signal length $n=64$ and the number of hidden frequencies $L=4$.}
		\label{fig:sim_MC_L4}
	\end{centering}
\end{figure}

\begin{figure}[!h]
	\begin{centering}
		\includegraphics[width=0.5\textwidth]{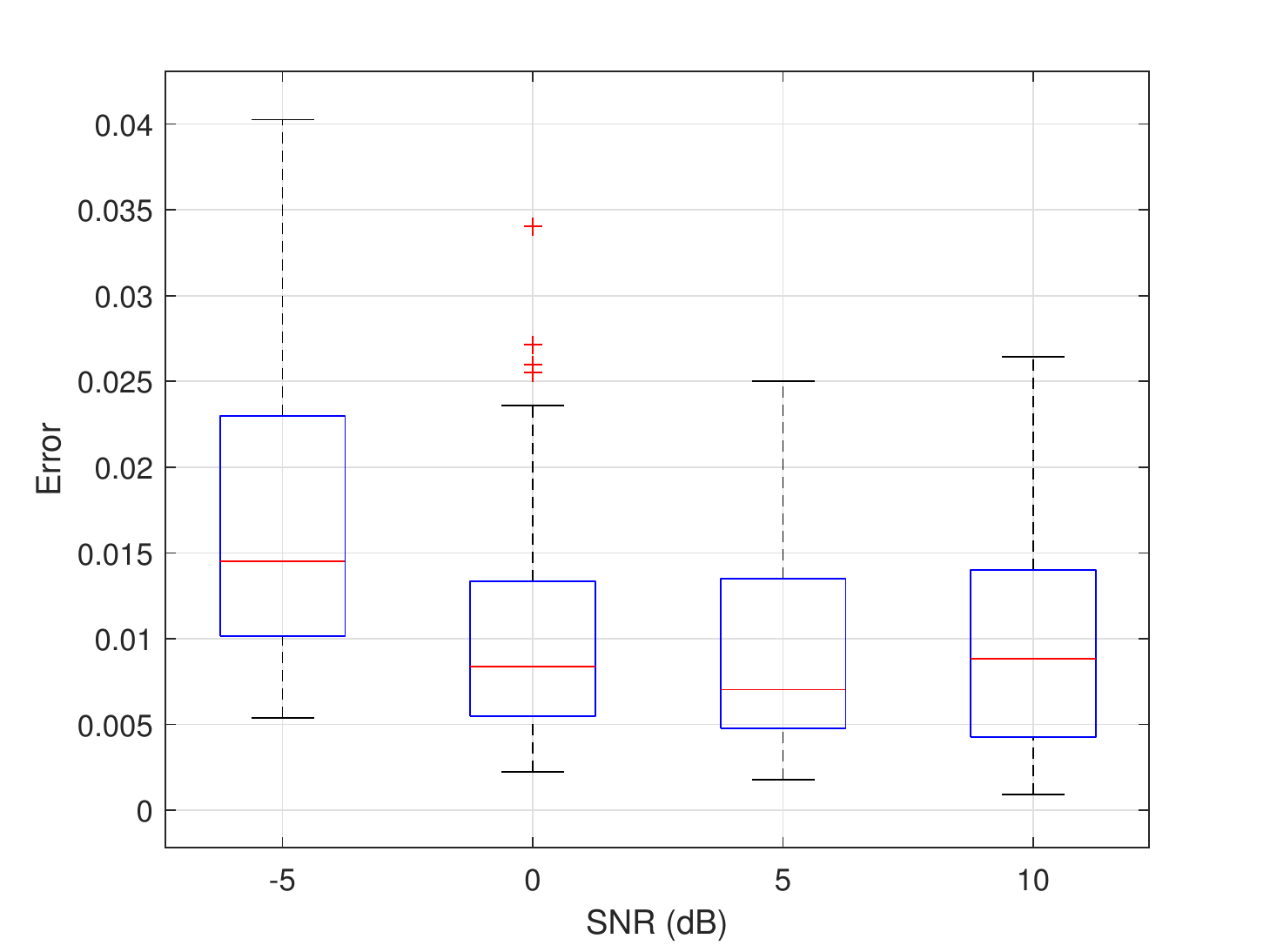}
		\caption{Errors of frequency estimation versus the SNR while the signal length $n=64$ and the number of hidden frequencies $L=8$.}
		\label{fig:sim_MC_L8}
	\end{centering}
\end{figure}

For the cases of $L=4$ and $L=8$, we further report the errors of frequency estimation in the $50$ successful trials as measured by the norm $\|\hat{\thetab}-\thetab\|$ using the boxplot in Figs.~\ref{fig:sim_MC_L4} and \ref{fig:sim_MC_L8}. It can be seen from Fig.~\ref{fig:sim_MC_L4} that the method enjoys a significant performance gain under large SNR, although such gain is less apparent in Fig.~\ref{fig:sim_MC_L8}. Moreover, a general picture is that the errors are quite small (up to the order of $10^{-2}$), meaning that the method is very robust against noise.

\section{Concluding remarks}
\label{sec:conclusions}

In this work, we have treated the problem of line spectrum representation for a given extendable covariance (multi-)sequence. We have shown the existence of such a representation using Carath\'{e}odory's theorem for convex hulls. We have also provided a sufficient condition for the uniqueness of the line spectrum representation for covariances on the boundary of the dual cone, and have demonstrated that in the special case of bivariate time series, the representation is indeed unique under a nondegeneracy condition. Equivalently, this leads to the Vandermonde decomposition for positive semidefinite singular block-Toeplitz matrices which finds application in frequency estimation using the atomic norm minimization approach. Given noiseless double-channel measurements, exact frequency recovery can be achieved via the solution of a convex optimization problem under a rank condition. The formulation can also be adapted to the noisy case. As revealed by numerical simulations, the method appears to work very well under various parameter configurations.

As for a possible future research direction, it seems interesting to investigate in which situation the rank condition in Theorem~\ref{thm_AN_noiseless} is automatically met. To this end, it may be helpful to further explore the connection between the alternative formulations involving \eqref{def_AN_xb} and \eqref{atomic_norm_Xb}.


\begin{ack}                               
The author would like to thank Prof.~Johan Karlsson for suggesting the formulation of Lemma~\ref{lem_prod_meas} which leads to the main result of Section~\ref{sec:existence}. The anonymous reviewers and the associate editor are also appreciated for their comments that helped to improve the quality of this paper. 
\end{ack}

\bibliographystyle{model4-names}
\bibliography{references}           

\appendix

\section{Proof of Theorem~\ref{thm_AN_noiseless}}

\begin{pf}
	We first prove the claim that $\|\xb\|_\Acal\geq p$. Let $\xb = \sum_{\ell} G(\theta_\ell) s_\ell$ be an atomic decomposition of $\xb$. Define the unit vector $u_\ell:=s_\ell/\|s_\ell\|$, the scalar $b:=\sum_{\ell}\|s_\ell\|$, and the Hermitian block-Toeplitz matrix $T(\Sigmab):=\sum_{\ell} \|s_\ell\| G(\theta_\ell) u_\ell u_\ell^* G^*(\theta_\ell)$. By construction, we have $\Sigma_0=\sum_{\ell} \|s_\ell\| u_\ell u_\ell^*$, and thus $\trace\Sigma_0=\sum_{\ell}\|s_\ell\|=b$. Moreover, we have
	\begin{equation}
	\bmat b&\xb^*\\\xb&T(\Sigmab) \emat = \sum_{\ell} \|s_\ell\| \bmat u_\ell^*\\ G(\theta_\ell) \emat u_\ell u_\ell^* \bmat u_\ell & G^*(\theta_\ell) \emat \geq 0.
	\end{equation}
	Therefore, $(b,\Sigmab)$ is a feasible point of the optimization problem \eqref{AN_semidef_program}, and by definition it holds that $p\leq\frac{1}{2}b + \frac{1}{2} \trace\Sigma_0 = \sum_{\ell}\|s_\ell\|$. Since the inequality holds for any atomic decomposition of $\xb$, it must hold for the infimum, i.e., $p\leq\|\xb\|_\Acal$.
	
	For the second claim, we need to show the inequality $\|\xb\|_\Acal\leq p$ under the additional rank condition. Due to the LMI constraint \eqref{LMI_constraint}, we have $T(\hat{\Sigmab})\geq0$, which plus nondegeneracy makes Corollary~\ref{cor_1d_unique} applicable. Thus we can write down the unique Vandermonde decomposition $T(\hat{\Sigmab}) = \sum_{\ell=1}^{\hat{L}} G(\hat{\theta}_\ell) \hat{Q}_\ell G^*(\hat{\theta}_\ell)$. By the theory of the generalized Schur complement \citep{zhang2006schur}, we have $\xb\in\range T(\hat{\Sigmab})$, which means that there exist vectors $\{\hat{s}_\ell\}$ such that
	\begin{equation}\label{atomic_decomp_xb}
	\xb = \sum_{\ell=1}^{\hat{L}} G(\hat{\theta}_\ell) \hat{s}_\ell =: \Gb(\hat{\thetab})\hat{\sbf}.
	\end{equation}
	The latter is a shorthand notation for the block-matrix-vector product in \eqref{y_meas_vec}. Again by the Schur complement, it holds that
	\begin{equation}\label{inequal_b}
	\begin{split}
	\hat{b} & \geq \xb^* T^\dagger(\hat{\Sigmab}) \xb \\
	& =\hat{\sbf}^* \Gb^*(\hat{\thetab}) \left[ \Gb(\hat{\thetab}) \hat{\Qb} \Gb^*(\hat{\thetab}) \right]^\dagger \Gb(\hat{\thetab})\hat{\sbf}
	\end{split}
	\end{equation}
	where $^\dagger$ denotes the Moore-Penrose pseudoinverse, and $\hat{\Qb}:=\diag\{\hat{Q}_1,\dots,\hat{Q}_{\hat{L}}\}$ is a block-diagonal matrix. Next we will simplify the above expression, during which the rank condition will play a role. Following the discussion after \eqref{decomp_Sigma_k}, we know $\hat{L}\leq\hat{r}\leq n+1$ so that the $2(n+1)\times2\hat{L}$ matrix $\Gb(\hat{\thetab})$ has linearly independent columns. Hence the matrix $\hat{\Qb}$ has rank $\hat{r}$. Consider also the eigen-decomposition $\hat{\Qb}=\hat{\Ub}\hat{\Lambdab}\hat{\Ub}^*$. The matrix $\hat{\Ub}$ has the shape
	\begin{equation}\label{struct_Ub}
	\bmat\hat{u}_{1,1}&\hat{u}_{1,2}&0&0&\cdots&0&0\\
	0&0&\hat{u}_{2,1}&\hat{u}_{2,2}&\cdots&0&0\\
	\vdots&\vdots&\vdots&\vdots&\ddots&\vdots&\vdots\\
	0&0&0&0&\cdots&\hat{u}_{\hat{L},1}&\hat{u}_{\hat{L},2}\emat
	\end{equation}
	where each pair $\hat{u}_{\ell,1},\,\hat{u}_{\ell,2}$ are orthonormal eigenvectors of the block $\hat{Q}_\ell$. Discarding the zero eigenvalues, we can write a ``thin'' decomposition $\hat{\Qb}=\hat{\Ub}_\trm \hat{\Lambdab}_\trm \hat{\Ub}_\trm^*$ where $\hat{\Lambdab}_\trm=\diag\{\hat{\lambda}_1,\dots,\hat{\lambda}_{\hat{r}}\}>0$. Taking a closer look at the representation \eqref{atomic_decomp_xb}, we can conclude that $\hat{\sbf}=\hat{\Ub}_\trm \bar{s}$ for some $\bar{s}\in\Cbb^{\hat{r}}$. Due to the special structure of $\hat{\Ub}_\trm$ derived from \eqref{struct_Ub}, it follows that each $\hat{s}_\ell$ is either the linear combination of $\hat{u}_{\ell,1},\,\hat{u}_{\ell,2}$ or proportional to one of the two. Both cases imply the inequality
	\begin{equation}\label{inequal_norm_s}
	\sum_{\ell=1}^{\hat{L}}\|\hat{s}_\ell\|\leq\sum_{k=1}^{\hat{r}}|\bar{s}_k|.
	\end{equation} 
	We can now continue \eqref{inequal_b} as follows
	\begin{equation}
	\begin{split}
	\hat{b} & \geq \bar{s}^*\hat{\Ub}_\trm^*\Gb^*(\hat{\thetab}) \left[\hat{\Ub}_\trm^*\Gb^*(\hat{\thetab})\right]^\dagger \hat{\Lambdab}_\trm^{-1} \left[\Gb(\hat{\thetab})\hat{\Ub}_\trm\right]^\dagger \Gb(\hat{\thetab})\hat{\Ub}_\trm \bar{s} \\
	& = \bar{s}^* \hat{\Lambdab}_\trm^{-1} \bar{s} = \sum_{k=1}^{\hat{r}} \frac{|\bar{s}_k|^2}{\hat{\lambda}_k}.
	\end{split}
	\end{equation}
	Finally, we arrive at
	\begin{equation}\label{p_geq_x_A}
	\begin{split}
	p & =\frac{1}{2}\hat{b} + \frac{1}{2} \trace\hat{\Sigma}_0 \\
	& \geq \frac{1}{2} \sum_{k=1}^{\hat{r}} \left( \frac{|\bar{s}_k|^2}{\hat{\lambda}_k} + \hat{\lambda}_k \right) \\
	& \geq \sum_{k=1}^{\hat{r}} |\bar{s}_k| \geq \sum_{\ell=1}^{\hat{L}}\|\hat{s}_\ell\| \geq \|\xb\|_\Acal,
	\end{split}
	\end{equation}
	where we have used \eqref{inequal_norm_s} and the definition of the atomic norm.
\end{pf}

\begin{remark}\label{rem:rank}
	It is worth noting from the above proof that in order for $p=\|\xb\|_\Acal$ to hold, all the inequalities in \eqref{p_geq_x_A} must hold with equality, which means that $\hat{L}=\hat{r}$, $\hat{\lambda}_k=|\bar{s}_k|=\|\hat{s}_k\|$, and the atomic decomposition \eqref{atomic_decomp_xb} achieves the atomic norm. In particular, each block $\hat{Q}_\ell$ in $\hat{\Qb}$ must have rank $1$. Adopting the interpretation in Remark \ref{rem:interp_cov_mat}, this implies that the two measurement channels are linearly correlated, also called \emph{coherent} in DOA estimation.
\end{remark}

\end{document}